\documentclass[twocolumn,english]{revtex4}
\usepackage[T1]{fontenc}
\usepackage[latin9]{inputenc}
\usepackage{babel}
\usepackage{amsmath}
\usepackage{amssymb}
\usepackage{graphicx}
\usepackage{esint}
\usepackage[unicode=true,
 bookmarks=true,bookmarksnumbered=true,bookmarksopen=true,bookmarksopenlevel=1,
 breaklinks=false,pdfborder={0 0 0},backref=false,colorlinks=false]
 {hyperref}
\hypersetup{pdftitle={Your Title},
 pdfauthor={Your Name},
 pdfpagelayout=OneColumn,pdfnewwindow=true,pdfstartview=XYZ,plainpages=false}
\usepackage{breakurl}

\makeatletter
\@ifundefined{textcolor}{}
{%
 \definecolor{BLACK}{gray}{0}
 \definecolor{WHITE}{gray}{1}
 \definecolor{RED}{rgb}{1,0,0}
 \definecolor{GREEN}{rgb}{0,1,0}
 \definecolor{BLUE}{rgb}{0,0,1}
 \definecolor{CYAN}{cmyk}{1,0,0,0}
 \definecolor{MAGENTA}{cmyk}{0,1,0,0}
 \definecolor{YELLOW}{cmyk}{0,0,1,0}
 }

\usepackage{babel}

\makeatother

\begin{document}

\title{Nonlinear Elimination of Spin-Exchange Relaxation of High Magnetic
Moments}

\author{Or Katz$^{1}$$^{3}$, Mark Dikopoltsev$^{2}$$^{3}$, Or Peleg$^{3}$,
Moshe Shuker$^{2}$$^{3}$, Jeff Steinhauer$^{2}$, Nadav Katz$^{1}$}

\affiliation{$^{1}$Racah Institute of Physics, The Hebrew University of Jerusalem,
Jerusalem 91904, Israel}

\affiliation{$^{2}$Department of Physics, Technion, Israel Institute of Technology,
Haifa, 32000, Israel}

\affiliation{$^{3}$Rafael Ltd, IL-31021 Haifa, Israel}
\begin{abstract}
Relaxation of the Larmor magnetic moment by spin-exchange collisions
has been shown to diminish for high alkali densities, resulting from
the linear part of the collisional interaction. In contrast, we demonstrate
both experimentally and theoretically the elimination of spin-exchange
relaxation of high magnetic moments (birefringence) in alkali vapor.
This elimination originates from the nonlinear part of the spin-exchange
interaction, as a scattering process of the Larmor magnetic moment.
We find counter-intuitively that the threshold magnetic field is the
same as in the Larmor case, despite the fact that the precession frequency
is twice as large.
\end{abstract}


\maketitle
Atomic relaxation mechanisms in vapor physics limit the precision
and performance of many spectroscopic measurements. For example, the
performance of atomic clocks \cite{atomic clocks} and sensitive magnetometers
\cite{magnetometers} is determined by these relaxations. The dominant
ground state relaxation mechanism is induced by spin-exchange collisions
of alkali atoms \cite{PURCELL - spin ex.}\cite{WITTKE spin ex.}.
However, it was found that at high atomic densities the Larmor (dipolar)
coherence, associated with any two nearest Zeeman splitted levels,
improves significantly with its frequency being slowed down \cite{happer spin ex experimental}\cite{happer 1977}.
This unique effect is known as Spin-Exchange Relaxation Free (SERF)
since the relaxation due to spin-exchange collisions is completely
suppressed and only next-order relaxation processes, such as the power
broadening of a pumping laser, become apparent. This effect sets the
stage for ultrasensitive vapor based magnetometery schemes \cite{Romalis SERF magnetometer}.

Since alkali vapor is essentially a multilevel quantum system, in
many coherent phenomena various orders of atomic coherence are involved,
each describing the coupling strength between one energy level to
another. The use of high orders of coherence introduced phenomena
such as nonlinear magneto-optical rotation \cite{NMOR} and had a
great contribution in improving magnetometery schemes \cite{NMOR - magnetometery}.
Although the high orders of coherence are also relaxed by spin-exchange
collisions, up to this point only the lowest order coherence (dipolar)
was studied experimentally and theoretically at high atomic densities
while the higher orders of coherence in this regime were completely
avoided.

In this Letter we demonstrate experimentally that the birefringent
coherence, associated with any two next-nearest Zeeman splitted levels,
also experiences SERF in the low magnetic field regime. We further
explain both numerically and analytically the origin of this effect
and deduce the main characteristics of this phenomena such as its
decoherence rate, oscillation frequency, and the magnetic field threshold
of the SERF regime. Magnetic moment orders higher than the birenfringence
are also treated, exhibiting a cascaded process leading to SERF. Finally,
we show that ultra-sensitive magnetometers, based on the birefringent
coherence, have twice the bandwidth of the Larmor magnetometer with
the same sensitivity.

\begin{figure}[tbh]
\begin{centering}
\includegraphics[width=7.5cm]{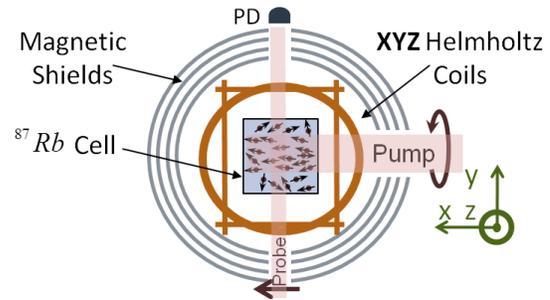} 
\par\end{centering}

\caption{{\small Experimental setup for the birefringent measurement. The cell
is initially polarized by a $\sigma_{+}$ pumping beam. Then, the
absorption of a $\pi$ probe is measured at a free induction decay
of the vapor in the presence of an applied magnetic field $B_{z}$.
\label{fig:Experimental-setup}}}
\end{figure}

To measure the Larmor and birefringent coherences a pump-probe scheme
is utilized (Fig. \ref{fig:Experimental-setup}), where the pump polarizes
the atomic vapor, and the absorption of the probe by the atoms is
measured. This absorption is described by the susceptibility of the
medium, which is in turn constructed by means of the different magnetic
moments of the atomic density matrix. A proper measurement of the
different moments is designed by choosing the relevant magnetic orders
of the susceptibility. 

The different orders of ground state coherence can be described by
expanding the ground state density matrix $\rho$ to its different
multiplets $\rho_{LM}\left(FF'\right)$ \cite{happer 1972}. Each
multiplet describes the coherence between the hyperfine levels $F$
and $F'$ with $L$ polarization moment distribution. The lowest three
multiplets are the isotropic $\left(L=0\right)$, dipolar $\left(L=1\right)$
and birefringent $\left(L=2\right)$ multiplets describing the population
of the hyperfine levels $F$,$F'$, the dipole moments, and the quadrupole
moments of their Zeeman sublevels, respectively. The $M$ quantum
number denotes the energy level spacing of the Zeeman splitting $M\hbar\omega_{B}$,
and is associated with the natural frequency $M\omega_{B}$. This
expansion is of special significance since single photon interactions,
being described by the complex susceptibility observable $\chi$ ,
involve solely the lowest three atomic multiplets \cite{suceptability - happer}.
All coherences satisfy $-L\leq M\leq L$ and we identify the $L=\left|M\right|=1$
as the Larmor coherence and $L=\left|M\right|=2$ as the birefringent
one. These coherences can be measured by monitoring $I_{l}$, the
intensity of a weak probe light beam after it propagates a distance
$l$ of alkali vapor \cite{suceptability - happer}
\begin{equation}
I_{l}=I_{0}\cdot\mbox{exp}\left[-2\pi l\sum_{ij}e_{i}^{*}\mbox{Im}\left(\left\langle \chi_{ij}\right\rangle \right)e_{j}\right]\label{eq:absorption}
\end{equation}
where $I_{0}$ is the probe intensity, $e_{i}$ is the electric polarization
vector of the probe, with: 
\begin{equation}
\left\langle \chi_{ij}\right\rangle =\sum_{LMFF'}\chi_{LMFF'}^{ij}\rho_{LM}\left(FF'\right).\label{eq:chi defenition}
\end{equation}

Here $\mbox{i,j}$ are the spatial Cartesian coordinates and $\chi_{LMFF'}^{ij}$
are the relative strengths coefficients \cite{suceptability - happer}\cite{happer 1967}.
From Eqs. \ref{eq:absorption},\ref{eq:chi defenition} the absorption
of a linear probe decays exponentially in time, with a decay rate
proportional to the birefringent multiplet $\rho_{22}$.

We measure the birefringent coherence by monitoring the absorption
of a linear probe and the Larmor coherence with a circular probe.
We use a $17^{3}\,\mbox{\ensuremath{\mbox{mm}^{3}}}$ cubic cell with
$\mbox{\ensuremath{^{87}}Rb}$, heated to temperature $T=111.5^{\circ}$
(atomic density $\sim10^{13}\,\mbox{c\ensuremath{m^{-3}}}$). We reduce
the collision rate with the cell walls by introducing $90\,\mbox{torr}$
of $N_{2}$ buffer gas. We control the applied magnetic field by shielding
the cell with $\mu\mbox{-metal}$ cylinders and three perpendicular
Helmholtz coils. A collimated distributed feedback laser beam with
a wide Gaussian profile ($23\,\mbox{mm}$ width) optically pumps the
atoms with circularly polarized light and effective pumping power
($0.97\,\mbox{mW}$) much lower than the saturation intensity. Then,
the pumping beam is switched off and a magnetic field $B\hat{z}$
is applied. As a result, the alkali spins start to precess around
the magnetic field. This precession decays due to the different relaxation
mechanisms. We probe this precession by measuring the absorption of
a weak narrow beam ($3\,\mbox{mm}$ width with a power of $2.5\,\mbox{\ensuremath{\mu}W}$).
The Larmor coherence is probed in the $F_{g}=2$ to $F_{e}=1$ $D1$
transition. To minimize the isotropic ($\mbox{"}T_{1}\mbox{"}$) probed
part from the birefringent part we $1\,\mbox{GHz}$ red-detune the
probe. The pressure of the buffer gas was chosen  such that the excited-state
hyperfine levels are still resolved, and the birefringent strengths
$\chi_{2MFF}^{ij}$ are observable \cite{budker resolveness}.
\begin{figure}[tbh]
\begin{centering}
\includegraphics[clip,width=8.6cm,height=4.8cm]{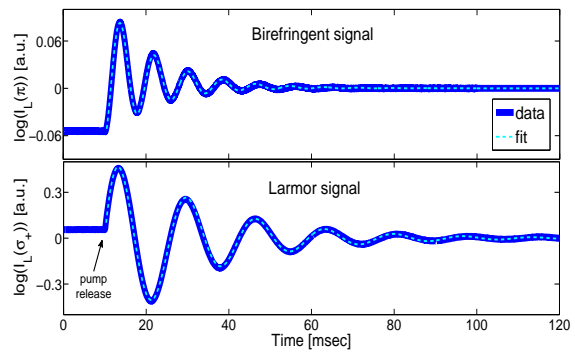} 
\par\end{centering}

\caption{{\small A typical measurement of the oscillation and decay of the
Larmor and birefringent coherences ($500$ averages). The birefringent
signal oscillates at twice the Larmor frequency and decays twice as
fast. The fit to each of the signals is shown (dotted curve).\label{fig:typical oscillation}}}
\end{figure}

A typical $\pi$ absorption (birefringent) measurement is shown in
Fig. \ref{fig:typical oscillation} for an applied magnetic field
of $28\,\mbox{nT}$ in the low magnetic field regime. A measurement
of the Larmor absorption for the same parameters is shown for reference.
The measured birefringent signal oscillates with frequency $\omega^{br}$
which is twice the frequency of the measured Larmor signal $\omega^{lr}$.
The decoherence rates and frequencies were determined by fitting each
signal to the simple model $f=Ae^{-\frac{t}{T_{1}}}+Ce^{-\Gamma_{0}t}\cos\left(\omega_{0}t\right)$,
assuming that both $\omega_{0}$ and $\Gamma_{0}$ are time independent
(a negligible dependence less than $8\%$ was observed). The fit was
performed for times longer than $t_{0}=300\,\mbox{\ensuremath{\mu}sec}$
satisfying $t_{0}\gg R_{SE}^{-1}$, where $R_{SE}\approx100\,\mbox{\ensuremath{\mu}sec}$
is the mean spin-exchange rate \cite{happer 1977}, to eliminate other
rapidly decaying moments. 

\begin{figure}[tbh]
\begin{centering}
\includegraphics[width=8.6cm]{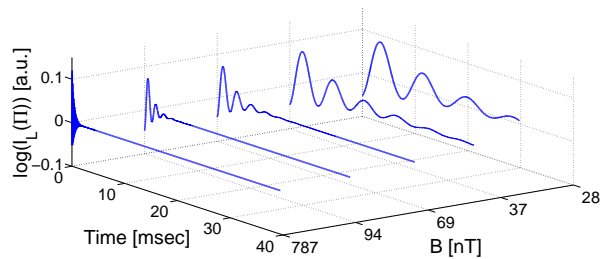} 
\par\end{centering}

\caption{{\small Measurement of the transition into the SERF regime of the
birefringent coherence. The decoherence rate decreases significantly
by decreasing the applied magnetic field. \label{fig: 3d graph}}}
\end{figure}

\begin{figure}[tbh]
\begin{centering}
\includegraphics[width=8.6cm]{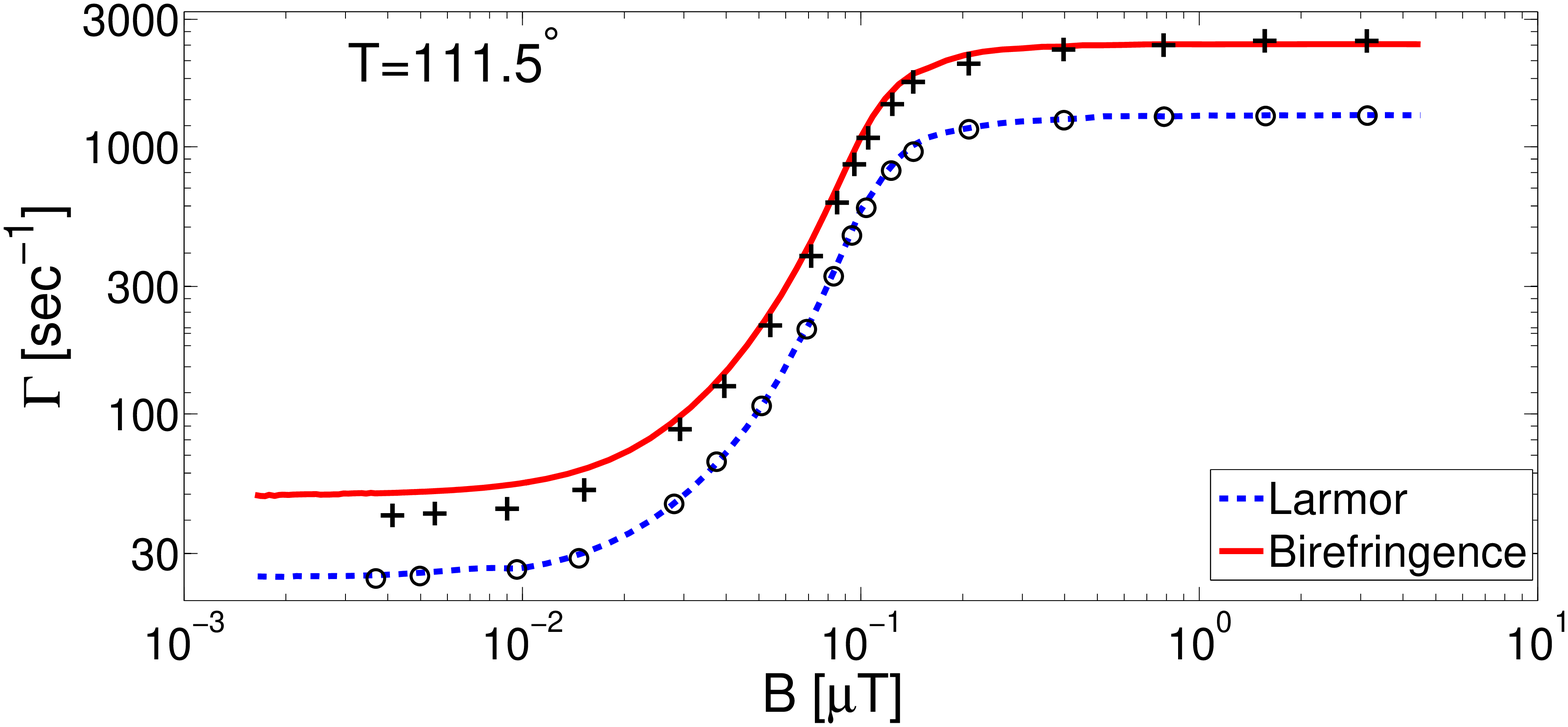} 
\par\end{centering}

\caption{{\small The measured magnetic field dependence of the decoherence
rates $\Gamma_{0}^{lr}\left(B\right)$(circle) and $\Gamma_{0}^{br}\left(B\right)$
(plus). It is apparent that both the Larmor and birefringent signals
experience SERF at the same magnetic field threshold. The birefringent
decoherence rate is twice the Larmor decay rate. The simulated decoherence
rates of the Larmor (dashed) and the birefringent (solid) coherences
are shown. \label{fig: decaying main graph}}}
\end{figure}

We find that the birefringent decoherence rate $\Gamma_{0}^{br}\left(B\right)$
decreases significantly by the decrease of the magnetic field as shown
qualitatively in Fig. \ref{fig: 3d graph} and quantitatively in Fig.
\ref{fig: decaying main graph}. 
 These graphs indicate the effect of rapid spin-exchange collisions
on both the Larmor and the birefringent coherences. At high magnetic
fields, the decoherence rate is limited by the high spin-exchange
collision rate $R_{SE}$ and approaches a constant value. By decreasing
the magnetic field, the decoherence rate associated with spin-exchange
collisions decreases, approaching quadratically to the lower plateau
in the low magnetic field regime. The lower plateau is determined
by other polarization ($\mbox{"}T_{1}\mbox{"}$) decoherence processes
such as the diffusion time of the atoms to the cell walls. In that
sense, the decoherence rate is now $T_{1}$ limited and the spin-exchange
relaxation is eliminated. In this regime it is found that $\Gamma_{0}^{br}\left(B\right)$
satisfies 
\begin{equation}
\Gamma_{0}^{br}\left(B\right)\cong2\Gamma_{0}^{lr}\left(B\right)\label{eq: twice the decaying rate}
\end{equation}
yielding a scaled SERF narrowing for the birefringent decoherence
rate $\Gamma_{0}^{br}\left(B\right)$ relative to the Larmor decoherence
rate $\Gamma_{0}^{lr}\left(B\right)$. Since the measured birefringent
signal is associated with the second harmonic of the Larmor frequency,
we identify this process as a nonlinear one \cite{Nonlinearity - boyed}
and will later explain its nature. We thus conclude that the birefringent
coherence experiences the nonlinear SERF effect at the same magnetic
field threshold of the Larmor coherence with its frequency and decoherence
rate doubled. 

To verify that the model of spin-exchange relaxations is valid for
the measurements, numerical ground state simulations are performed,
considering the following Liouville equation \cite{Romalis SERF magnetometer}
\begin{eqnarray}
\frac{d\rho}{dt} & = & A_{hfs}\frac{\left[\mathbf{I\cdot S},\rho\right]}{i\hbar}+\omega_{B}\frac{\left[S_{z},\rho\right]}{i\hbar}+R_{SE}\left(4\alpha\mathbf{S}\left\langle \mathbf{S}\right\rangle -\mathbf{A}\cdot\mathbf{S}\right)\nonumber \\
 &  & -R_{SD}\mathbf{A}\cdot\mathbf{S}\label{eq:simulation equation}
\end{eqnarray}
where $\mathbf{I}$ and $\mathbf{S}$ are the nuclear and electronic
spin observables, $A_{hfs}$ is the hyperfine coupling constant of
$\mbox{\ensuremath{^{87}}Rb}$, $\omega_{B}$ is the bare electron
precession frequency and $R_{SD}$ is the spin-destructing collision
rate. The operators $\alpha$ and $\mathbf{A}$ are the nuclear and
electronic parts of the density matrix $\rho=\alpha+\mathbf{A\cdot S}$
\cite{happer 1977}. Hyperfine coherences ($F\neq F'$) are neglected
to improve the simulation runtime \cite{SERF high polarization-1}. 

The predictions of the numerical model are plotted in Fig. \ref{fig: decaying main graph}.
We used a single fitting parameter ($R_{SD}=147\,\mbox{se\ensuremath{c^{-1}}}$),
determined by fitting the decoherence rate of the Larmor coherence
at low magnetic fields. The birefringence is plotted using no free
parameters. The simulations also yield the slight time dependence
of the measured frequencies and decoherence rates, resulting from
the relaxation of the mean spin polarization. Therefore, the subscript
$\mbox{"}0\mbox{"}$ denotes the values of the decoherence rates {\small $\Gamma_{0}^{lr}$},
{\small $\Gamma_{0}^{br}$} in the low-polarization regime. Evidently,
the simulations verify that all the processes governing the birefringent
coherence are included in Eq. \ref{eq:simulation equation}. Furthermore,
the main observations of the SERF magnetic field threshold, the decoherence
rate and precession frequency are well described by the simulations.

These results are surprising in two aspects. Firstly, one would expect
the birefringent coherence to experience SERF at magnetic fields satisfying
$\omega_{0}^{br}=2\omega_{0}^{lr}\lesssim R_{SE}$ while de facto
it satisfies the same condition as the Larmor coherence $\omega_{0}^{lr}\lesssim R_{SE}$.
Secondly, Eq. \ref{eq: twice the decaying rate} is rather surprising
since neither a spin vector model \cite{SERF high polarization-1}
nor a motional narrowing-based model \cite{happer 1977} can describe
it. These models are based on the linearized spin-exchange interaction,
neglecting nonlinear terms arising from the interaction with the mean
spin polarization. In the following, we show that the dynamics of
the birefringent coherence relies entirely on this nonlinear mechanism.
To examine this nonlinearity using perturbation analysis, we rephrase
Eq. \ref{eq:simulation equation} using the super-operator formalism
\cite{happer 1977}

\begin{equation}
\frac{d\rho}{dt}=\left(W+Z+E+Q\left(\rho\right)\right)\rho\label{eq:main eq dynamics}
\end{equation}
where $W$ denotes the hyperfine interaction, $Z$ denotes the interaction
of the external magnetic field with the electron spin, and $E$ and
$Q\left(\rho\right)$ denote the linear and nonlinear terms of the
spin-exchange relaxation. For simplicity we neglect the spin destruction
interaction. To describe Eq. \ref{eq:main eq dynamics} in the interaction
picture we use the basis of eigenoperators $\left|LM\pm\right\rangle $
(associated with the magnetic $L,M$ multiplets) of the linear super-operator
$W+Z+E$. These eigenoperators were first calculated in \cite{happer 1977}
for the multiplets $L=0,1$. The calculation is extended for $L>1$
in \cite{supplemnetary}. In the linear theory, $\rho$ satisfies
\begin{equation}
\rho\left(t\right)|^{linear}=\sum_{LM\pm}\rho_{LM\pm}\left(0\right)\cdot e^{\lambda_{\pm}^{LM}t}\left|LM\pm\right\rangle 
\end{equation}
where the magnetic eigenvalues $\lambda_{\pm}^{LM}$ describe the
dynamics of $\rho$ completely. In the low magnetic field regime ($\omega_{0}\lesssim R_{SE}$)
these eigenvalues are given by \cite{supplemnetary}
\begin{equation}
\lambda_{\pm}^{LM}\approx\pm\frac{ib_{L}M}{2c_{L}}\omega_{0}-R_{SE}\left(a_{L}\mp c_{L}\right)-\mathcal{O}\left(\frac{\omega_{0}^{2}}{R_{SE}}\right)\label{eq:SERF eigenvalues linear-1}
\end{equation}
where $\omega_{0}=\omega_{B}/\left(2I+1\right)$ is the slowed down
Larmor frequency and the coefficients $a_{L},b_{L},c_{L}$ are given
in \cite{supplemnetary}.

Using the $\left|LM\pm\right\rangle $ basis, we transform Eq. \ref{eq:main eq dynamics}
to the interaction picture to include the nonlinear interaction $Q\left(\rho\right)\rho$

\begin{eqnarray}
\rho_{LM\pm}\left(t\right) & = & \rho_{LM\pm}\left(0\right)e^{\lambda_{\pm}^{LM}t}+2R_{SE}\sum_{l,m,k,\pm'}Q\{k\}_{lm\pm'}^{LM\pm}\nonumber \\
 & \times & \intop e^{\lambda_{\pm}^{LM}\left(t-t'\right)}\rho_{lm\pm'}\left(t'\right)\left\langle S_{k}\right\rangle dt'\label{eq: full integral equation}
\end{eqnarray}
where the super-operator $Q$ is given in the magnetic multiplets
representation $\left|LM\pm\right\rangle $ by %
\footnote{The neglect of the hyperfine multiplets is discussed in \cite{supplemnetary}. %
}

\begin{equation}
\left\langle LM\pm\right|Q\left|lm\pm'\right\rangle =2R_{SE}\sum_{k=-1}^{1}Q\left\{ k\right\} {}_{lm\pm'}^{LM\pm}\left\langle S_{k}\right\rangle .\label{eq:Q definition}
\end{equation}
 Here $k$ denotes spherical coordinates and the coefficients $Q\left\{ k\right\} {}_{lm\pm'}^{LM\pm}$
are defined by angular momentum transformations in \cite{supplemnetary}.
We note that $Q\left\{ k\right\} {}_{lm\pm'}^{LM\pm}$ satisfy the
selection rules implied by the condition

\begin{equation}
Q\left\{ k\right\} {}_{lm\pm'}^{LM\pm}\propto C\left(l,1;mk;LM\right)\left(1-\delta_{l0}\right)\label{eq: Q coeff}
\end{equation}
where $C$ is the Clebsch-Gordan coefficient. By Eq. \ref{eq:Q definition},
the typical rate of $Q\left\{ k\right\} $ is $2R_{SE}\left\langle S_{k}\right\rangle $.
Since this rate satisfies $2R_{SE}\left\langle S_{k}\right\rangle \leq R_{SE}$,
in the low-polarization regime, one can approximate the contribution
of $Q$ by using a time-dependent perturbation theory. This method
is applied by approximating $Q\left(\rho\right)\approx Q\left(\rho|^{linear}\right)$
in Eq. \ref{eq: full integral equation}. In the low magnetic field
regime of the linear theory (Eq. \ref{eq:SERF eigenvalues linear-1}),
the only non vanishing components of $\rho_{lm\pm}$ for long times
$t\gg R_{SE}^{-1}$ are 
\begin{equation}
\rho_{lm\pm}\left(t\right)|_{t\gg R_{SE}^{-1}}^{linear}\rightarrow\rho_{lm+}\left(t_{0}\right)e^{\lambda_{+}^{lm}\left(t-t_{0}\right)}\left(\delta_{l0}+\delta_{l1}\right)\label{eq: linear long time approximation}
\end{equation}
where $t_{0}$ is some initial time satisfying $t_{0}\gg R_{SE}^{-1}$.
Applying the last equation to $\left\langle S_{k}\right\rangle $
one obtains 
\begin{equation}
\left\langle S_{k}\left(t\right)\right\rangle |_{t\gg R_{SE}^{-1}}^{linear}\rightarrow\left\langle S_{k}\left(t_{0}\right)\right\rangle e^{\lambda_{+}^{1k}\left(t-t_{0}\right)}.\label{eq: S expectaion linear}
\end{equation}

The $L\neq1$ Zeeman multiplets are given by substituting the linear
solutions (Eqs. \ref{eq: linear long time approximation},\ref{eq: S expectaion linear})
in Eq. \ref{eq: full integral equation}. Performing integration for
times $t\gg R_{SE}^{-1}$ yields 
\begin{eqnarray}
\rho_{LM\pm}\left(t\right) & \approx & \sum_{m,k=-1}^{1}\frac{2R_{SE}\left\langle S_{k}\left(t_{0}\right)\right\rangle \rho_{1m+}\left(t_{0}\right)}{-\lambda_{\pm}^{LM}+\lambda_{+}^{1m}+\lambda_{+}^{1k}}\nonumber \\
 & \times & Q\{k\}_{1m+}^{LM\pm}e^{\left(\lambda_{+}^{1m}+\lambda_{+}^{1k}\right)\left(t-t_{0}\right)}\label{eq: non linear apprx general}
\end{eqnarray}
where rapid decaying terms (with typical decoherence rate of $\Gamma\approx R_{SE}$)
are neglected. Considering the Clebsch-Gordan coefficient in Eq. \ref{eq: Q coeff},
the selection rule $k+m=M$ is obtained. Introducing this condition
to the birefringent multiplets $L=\left|M\right|=2$ in Eq. \ref{eq: non linear apprx general}
one obtains 

\begin{equation}
\rho_{2M\pm}\left(t\right)\approx\left(\frac{2\left\langle S_{k}\left(t_{0}\right)\right\rangle \rho_{1k+}\left(t_{0}\right)}{a_{2}\mp c_{2}}Q\{k\}_{1k+}^{2M\pm}\right)e^{2\lambda_{+}^{1k}\left(t-t_{0}\right)}\label{eq:sacttering amplitude}
\end{equation}
for $k=M/2$. Thus, the evolution of the birefringent multiplets is
determined by a single exponent. We can now identify the new birefringent
eigenvalue $\lambda^{2M}|_{M=\pm2}^{nonlinear}\equiv\pm i\omega_{0}^{br}-\Gamma_{0}^{br}$
with 
\begin{equation}
\lambda^{2M}|_{M=\pm2}^{nonlinear}=2\lambda_{+}^{1M}|_{M=\pm1}^{linear}.\label{eq:new eigenvalue}
\end{equation}

Since the Larmor multiplets with eigenvalues $\lambda_{+}^{1k}|_{k=\pm1}^{linear}=\pm i\omega_{0}^{lr}-\Gamma_{0}^{lr}$
experience SERF, we find that the birefringent multiplets also experience
SERF with both the frequency and the decoherence rate doubled, in
complete correspondence with the experiment and simulations. However,
the mechanism for this spin-exchange relaxation elimination is completely
different from the elimination introduced by the linear theory. The
emergence of the birefringent $M=\pm2$ coherence can be interpreted
as an induced nonlinear scattering of the Larmor coherence ($\rho_{1k+}$)
by the effective mean field potential $Q\left(\left\langle S_{k}\right\rangle \right)|^{linear}$
(Eq. \ref{eq: full integral equation}). Both the Larmor coherence
and the mean electronic spin $\left\langle \mathbf{S}\right\rangle $
experience the linear SERF effect but their coupling induces the birefringence.
Therefore, the birefringent SERF magnetic threshold $R\gtrsim\omega_{0}^{lr}$
is the same as the Larmor threshold. Moreover, the observed birefringent
doubled slowed-down frequency results from the time dependence of
the scattering rate ($\left\langle \mathbf{S}\left(t\right)\right\rangle $)
and not from a statistical mixture of birefringent precession of different
hyperfine levels. This is the reason why this effect can not be explained
in terms of the linear model and is therefore designated as the nonlinear
SERF effect.

Although resulting from a nonlinear interaction, it should be noted
that these eigenvalues are not $\rho$ or $\left\langle S_{k}\right\rangle $
dependent, since $\left\langle S_{k}\right\rangle $ is mainly dominated
by $\left\langle S_{k}\left(t\right)\right\rangle |^{linear}$. However,
by considering higher order terms in the perturbation expansion (Eq.
\ref{eq: full integral equation}) beyond the low-polarization regime,
polarization-dependent corrections of the slowing down factor $q\left(P\right)$
can be obtained %
\footnote{This dependence was demonstrated for the Larmor coherence in \cite{SERF high polarization-1}.%
}. Furthermore, these higher order perturbation terms induce the scattering
of higher order moments ($L>2)$. Following the selection rules of
$Q$ in Eq. \ref{eq: Q coeff}, any induced moment $L$ is scattered
only by a lower moment $L-1$. Thus, by iterating the perturbative
analysis, one can show that these higher moments follow a cascade
of nonlinear SERF scattering. The measurement of these higher moments
is more complicated \cite{budker resolveness}, and will be treated
in a future publication. 

For a given magnetic field, the birefringent coherence precesses twice
as fast as the Larmor coherence. Thus, a birefringent based magnetometer
will have twice the bandwidth of a Larmor-based magnetometer. The
fundamental sensitivity \cite{Romalis SERF magnetometer} ratio between
a birefringent based magnetometer and a Larmor-based magnetometer
would be $\sqrt{1/\left(2\eta_{br}\right)}$ where $\eta_{br}=N_{br}/N_{lr}$
is the relative birefringent number of oscillators with respect to
the Larmor number of oscillators. This can be approximated by $\eta_{br}\approx\left|\rho_{22+}/\rho_{11+}\right|$.
Using Eq. \ref{eq:sacttering amplitude}, one can estimate $\eta_{br}\approx1$
for moderate polarizations. Thus, the fundamental sensitivity of a
birefringent based magnetometer will have the same sensitivity as
a Larmor-based magnetometer.

In conclusion, we have demonstrated experimentally and theoretically
that the birefringent coherence experiences nonlinear SERF at low
magnetic fields, with the same magnetic field threshold but with doubled
frequency and decoherence rate. We have shown that the birefringence
coherence originates from a nonlinear scattering of the Larmor coherence
by the mean electronic spin of the vapor, and that other higher orders
of coherence experience a similar scattering process . Finally, we
have shown that birefringent based magnetometers will have the same
sensitivity as a Larmor based magnetometers with twice the bandwidth.

\end{document}


\title{Supplementary Material for the article in Physics Review Letters
: \medskip{}
Nonlinear Elimination of Spin-Exchange Relaxation of High Magnetic
Moments}

\author{Or Katz, Mark Dikopoltsev, Or Peleg, Moshe Shuker, Jeff Steinhauer,
Nadav Katz}

\affiliation{}

\maketitle
We present a comprehensive description of the linear spin-exchange
dynamics, extending the work of \cite{happer 1977} for any magnetic
multiplet $L$. The linear theory is the unperturbed (zeroth order)
solution of the nonlinear treatment in the main text. We then derive
the exact form of the coefficients $Q\left\{ k\right\} {}_{lm\pm'}^{LM\pm}$
associated with the quadratic spin-exchange interaction.

The linear eigenvalues (in the absence of the nonlinear interaction
$Q\rho$) are found by solving the linear equation (derived from Eq.
$5$ of the main text)
\begin{equation}
\frac{d\rho}{dt}=\left(W+Z+E\right)\rho\label{eq:linear dynamics}
\end{equation}
where the super-operators are the hyperfine structure interaction
($W$), the external magnetic field interaction with the electron
and nuclear spins ($Z$) and the linear spin exchange interaction
term $(E$). Since the hyperfine coupling is the dominant ground state
interaction, it will become fruitful to represent every operator $A$
with the super-operator $W$ eigenbasis $\left|LMFF'\right\rangle $
\begin{equation}
A=\sum_{LMFF'}A_{LM}\left(FF'\right)\left|LMFF'\right\rangle 
\end{equation}
where $\left|LMFF'\right\rangle $ denotes the coupled spherical tensor
$T_{LM}\left(FF'\right)$ as defined in \cite{happer 1972}. In this
representation , according to \cite{happer 1977}, the matrix elements
of $W$, $Z$ and $E$ of Eq. \ref{eq:linear dynamics} are given
by 

\begin{equation}
\left\langle LMFF'\right|W\left|lmff'\right\rangle =-i\omega_{hfs}\left(FF'\right)\delta_{Ll}\delta_{Mm}\delta_{Ff}\delta_{ff'}\label{eq:W definition}
\end{equation}

\begin{equation}
\left\langle LMFF'\right|Z\left|lmff'\right\rangle =-i\omega_{0}\left(F\right)M\delta_{Ll}\delta_{Mm}\delta_{FF'}\delta_{ff'}\delta_{Ff}\label{eq:Z definition}
\end{equation}

\begin{equation}
\begin{array}{ccc}
\left\langle LMFF'\right|E\left|lmff'\right\rangle  & = & R_{SE}\delta_{Ll}\delta_{Mm}\left\{ -\delta_{Ff}\delta_{F'f'}\right.\\
 & + & X\left(FF'L\right)X\left(ff'L\right)\\
 & + & \left.Y\left(FF'L\right)Y\left(ff'L\right)\delta_{L1}\right\} 
\end{array}\label{eq:E definition}
\end{equation}
where the hyperfine frequency $\omega_{hpf}\left(FF'\right)$, the
magnetic slowed down frequency $\omega_{0}$, and the coefficients
$X$ and $Y$ are given by

\[
\omega_{hpf}\left(FF'\right)=\frac{A_{hpf}}{2}\left(F\left(F+1\right)-F'\left(F'+1\right)\right)
\]

\begin{equation}
\omega_{0}\left(F\right)=\frac{g_{S}\mu_{B}}{\hbar[I]}\frac{F\left(F+1\right)+S\left(S+1\right)-I\left(I+1\right)}{2F\left(F+1\right)}B\label{eq: bohr frequency}
\end{equation}

\[
X\left(FF'L\right)=\sqrt{\frac{\left[F\right]\left[F'\right]}{\left[S\right]}}W_{racah}\left(F'SLI;IF\right)
\]

\[
Y\left(FF'L\right)=\sqrt{\frac{\left[F\right]\left[F'\right]}{\left[I\right]}}W_{racah}\left(IFSL;SF'\right)
\]
where $A_{hpf}$ is the hyperfine coupling coefficient, $B$ is the
magnetic field, $W_{racah}$ is the Racah's W coefficient and the
notation $\left[\cdot\right]$ is defined as $\left[A\right]\equiv2A+1$.

For small magnetic fields, the ground state hyperfine levels are well
resolved $\omega_{hfs}\ggg\omega_{0},R_{SE}$. Thus, one can approximate
the effect of $Z$ and $E$ in Eq. \ref{eq:linear dynamics} by using
perturbation methods on the eigenstates $\left|LMFF'\right\rangle $.
Between different hyperfine multiplets $F\neq F'$, the eigenvalues
of the hyperfine coherences are given by
\begin{equation}
\lambda_{hpf}\left(FF'\right)\approx i\omega_{hfs}\left(FF'\right)-R_{SE}.\label{eq:hyperfine eigenvalues}
\end{equation}
Therefore, the hyperfine decoherence rate is $\Gamma_{hfs}=R_{SE}$,
and in the regime of rapid spin exchange rate, the decaying of $\rho_{LMFF'}\left(t\right)$
for $F\neq F'$ is rapid.

Within the Zeeman multiplets $F=F'$, the eigenvalue zero (Eq. \ref{eq:W definition})
is associated with the degenerate eigenstates $\left|LMaa\right\rangle $
and $\left|LMbb\right\rangle $ where $a=I+1/2$ and $b=I-1/2$. Since
$E$ and $Z$ are $L,M$ invariant (Eqs. \ref{eq:Z definition},\ref{eq:E definition}),
it is sufficient to find the Zeeman eigenvalues by setting
\begin{equation}
\left|\begin{array}{cc}
\left\langle LMaa\right|E+Z\left|LMaa\right\rangle -\lambda & \left\langle LMaa\right|E+Z\left|LMbb\right\rangle \\
\left\langle LMbb\right|E+Z\left|LMaa\right\rangle  & \left\langle LMbb\right|E+Z\left|LMbb\right\rangle -\lambda
\end{array}\right|=0\label{eq:matrix diagonalization}
\end{equation}
with the aid of Eqs. \ref{eq:Z definition},\ref{eq:E definition},
the resulting eigenvalues of the Zeeman multiplets are 
\begin{equation}
\lambda_{\pm}^{LM}=-a_{L}R_{SE}\pm\sqrt{-M^{2}\omega_{0}^{2}+ib_{L}M\omega_{0}R_{SE}+c_{L}^{2}R_{SE}^{2}}\label{eq:linear eigenvalues}
\end{equation}
where the coefficients $a_{L}$, $b_{L}$ and $c_{L}$ for different
$L$ values are given in table \ref{tab: eigenvalues_coefficients}
for $\mbox{\ensuremath{^{87}}Rb}$ (with spin $I=3/2$). 
\begin{table}
\begin{centering}
\begin{tabular}{|c|c|c|c|c|c|}
\hline 
$L$ & $a_{L}$ & $b_{L}$ & $c_{L}$ & $\alpha_{L}$ & $\beta_{L}$\tabularnewline
\hline 
\hline 
$0$ & $\nicefrac{1}{2}$ & $\nicefrac{1}{4}$ & $\nicefrac{1}{2}$ & $\sqrt{\nicefrac{5}{8}}$ & $\sqrt{\nicefrac{3}{8}}$\tabularnewline
\hline 
$1$ & $\nicefrac{3}{8}$ & $\nicefrac{1}{2}$ & $\nicefrac{3}{8}$ & $\sqrt{\nicefrac{5}{6}}$ & $\sqrt{\nicefrac{1}{6}}$\tabularnewline
\hline 
$2$ & $\nicefrac{11}{16}$ & $\nicefrac{1}{4}$ & $\nicefrac{5}{16}$ & $\sqrt{\nicefrac{7}{10}}$ & $\sqrt{\nicefrac{3}{10}}$\tabularnewline
\hline 
$3$ & $\nicefrac{7}{8}$ & $\nicefrac{1}{4}$ & $\nicefrac{1}{8}$ & $1$ & $0$\tabularnewline
\hline 
$4$ & $1$ & $\nicefrac{1}{4}$ & $0$ & $1$ & $0$\tabularnewline
\hline 
\end{tabular}
\par\end{centering}

\caption{\label{tab: eigenvalues_coefficients} Coefficients for the eigenvalues
$\lambda_{\pm}^{LM}$ in Eq. \ref{eq:linear eigenvalues} for $\mbox{\ensuremath{^{87}}Rb}$.
The coefficients for $L=0,1$ were calculated in \cite{happer 1977}. }
\end{table}

In the low magnetic fields regime ($\omega_{0}\ll R_{SE}$), for $L\neq4$,
Eq. \ref{eq:linear eigenvalues} takes the simple form 
\begin{equation}
\lambda_{\pm}^{LM}\approx\pm\frac{ib_{L}M}{2c_{L}}\omega_{0}-R_{SE}\left(a_{L}\mp c_{L}\right)-\mathcal{O}\left(\frac{\omega_{0}^{2}}{R_{SE}}\right)\label{eq:SERF eigenvalues linear}
\end{equation}
where the imaginary part $\left(b_{L}M/2c_{L}\right)\omega_{0}$ is
the spin-exchange slowed down precession frequency while the real
part $R_{SE}\left(a_{L}\mp c_{L}\right)-\mathcal{O}\left(\omega_{0}^{2}/R_{SE}\right)$
is the multiplet damping rate. The linear eigenvalues of Eq. \ref{eq:SERF eigenvalues linear}
are associated with the linear eigenstates $\left|LM\pm\right\rangle $
of the linear Liouvillian (Eq. \ref{eq:linear dynamics}), given by
\begin{equation}
\left(\begin{array}{c}
\left|LM+\right\rangle \\
\left|LM-\right\rangle 
\end{array}\right)=\left(\begin{array}{cc}
\alpha_{L} & \beta_{L}\\
-\beta_{L} & \alpha_{L}
\end{array}\right)\left(\begin{array}{c}
\left|LMaa\right\rangle \\
\left|LMbb\right\rangle 
\end{array}\right)\label{eq: eigenstates decomposition}
\end{equation}
where the coefficients $\alpha_{L}$ and $\beta_{L}$ are also listed
in table \ref{tab: eigenvalues_coefficients} for the $\omega_{0}\ll R_{SE}$
regime. These new eigenstates represent a mixing between the two upper
and lower hyperfine levels $a$ and $b$, each precessing at a typical
rate $\omega_{0}\left(F\right)$ (Eq. \ref{eq: bohr frequency}).
The weighted net frequency of this coherent mixed state is $\left(b_{L}M/2c_{L}\right)\omega_{0}$.
In correspondence to Eq. \ref{eq: eigenstates decomposition}, we
define the unperturbed Zeeman multiplets by the mixing 
\[
\left(\begin{array}{c}
\rho_{LM+}\left(t\right)\\
\rho_{LM-}\left(t\right)
\end{array}\right)=\left(\begin{array}{cc}
\alpha_{L} & \beta_{L}\\
-\beta_{L} & \alpha_{L}
\end{array}\right)\left(\begin{array}{c}
\rho_{LMaa}\left(t\right)\\
\rho_{LMbb}\left(t\right)
\end{array}\right).
\]
Using Eq. \ref{eq:SERF eigenvalues linear}, for long times $t\gg R_{SE}^{-1}$,
the term $\rho_{LM-}\left(t\right)$ decays faster than $\rho_{LM+}\left(t\right)$.
Thus, one shall recognize the decoherence rate of the multiplet $\rho_{LMFF}\left(t\right)|_{t\gg R_{SE}^{-1}}$
by the rate $\Gamma_{L}=\mbox{Re}\left(-\lambda_{+}^{LM}\right)$.
In this regime, the first order of $\Gamma^{lr}$ ($L=1$) cancels
out (Eq. \ref{eq:SERF eigenvalues linear}), yielding the low linear
SERF decoherence rate proportional to $\omega_{0}^{2}/R_{SE}$. However
the resulting decoherence rate of other multiplets $\Gamma_{L}$ for
$L\neq1$ in Eq. \ref{eq:SERF eigenvalues linear} increases linearly
with $R_{SE}$ at low magnetic fields. The implied slowed down frequency
is also different than the measured birefringent frequency shown in
the main text. Thus, in the scope of the linear theory all other multiplets,
including the birefringent one $\rho_{2M+}$, do not experience linear
SERF, but decay quickly. 

According to \cite{happer 1977}, the spin-exchange interaction $Q$
is given in the uncoupled spherical tensors basis $\left|\Lambda\mu;lm\right\rangle $
by
\begin{equation}
Q\left|\Lambda\mu;lm\right\rangle =2R_{SE}\delta_{l0}\left(1-\delta_{\Lambda0}\right)\sum_{k=-1}^{1}\left\langle S_{k}\right\rangle \left|\Lambda\mu;1k\right\rangle .\label{eq:Q represntation uncoupled basis}
\end{equation}
This super-operator can be represented by the coupled spherical representation
$\left|LMFF'\right\rangle $ by considering the coupled to uncoupled
spherical tensors transformation
\begin{align*}
\left|LMFF'\right\rangle  & =\sum_{\Lambda\mu;lm}\sqrt{\left[F\right]\left[F'\right]\left[\Lambda\right]\left[l\right]}\left(\begin{array}{ccc}
I & I & j_{1}\\
S & S & j_{2}\\
F & F' & L
\end{array}\right)\\
\times & C\left(\Lambda,l;\mu,m,L,M\right)\left|\Lambda\mu;lm\right\rangle 
\end{align*}
where $\left(\cdot\right)$ denotes the Wigner 9-j symbol and $C$
the Clebsch-Gordan coefficient. Applying this transformation to Eq.
\ref{eq:Q represntation uncoupled basis} yields the coupled representation
\begin{equation}
\left\langle LMFF'\right|Q\left|lmff'\right\rangle =2R_{SE}\sum_{k=-1}^{1}Q\left\{ k\right\} {}_{lmff'}^{LMFF'}\left\langle S_{k}\right\rangle \label{eq:Q definition}
\end{equation}
where the coefficients $Q\{k\}_{lmff'}^{LMFF'}$ are given by

\begin{eqnarray*}
Q\left\{ k\right\} {}_{lmff'}^{LMFF'} & =\sum & \sqrt{\frac{3}{2}\left[f'\right]\left[f'\right]\left[F\right]\left[F'\right]\left[L\right]}W_{rach}\left(f'SlI;If\right)\\
 &  & \times C\left(l,1;mk;LM\right)\left(\begin{array}{ccc}
I & I & l\\
S & S & 1\\
F & F' & L
\end{array}\right)\left(1-\delta_{l0}\right)\\
 & \equiv & \left\langle LMFF'\right|Q\left\{ k\right\} \left|lmff'\right\rangle .
\end{eqnarray*}

We can further represent the magnetic parts (same hyperfine levels)
of this interaction by the linear eigenstates $\left|LM\pm\right\rangle $.
The coefficients $Q\left\{ k\right\} {}_{lm\pm'}^{LM\pm}\equiv\left\langle LM\pm\right|Q\left\{ k\right\} \left|lm\pm'\right\rangle $
are thus directly derived by applying the transformation of Eq. \ref{eq: eigenstates decomposition}
twice. Therefore, the eigenbasis coefficients $Q\left\{ k\right\} {}_{lm\pm'}^{LM\pm}$
are a superposition of the coupled coefficients $Q\left\{ k\right\} {}_{lmff'}^{LMFF'}$,
yielding the expression of $Q\left\{ k\right\} {}_{lm\pm'}^{LM\pm}$
in the main text.

We finally note that in the last transformation we considered the
coupling of the $L,M$ and $l,m$ multiplets only within the same
hyperfine levels. However, the interaction $Q$ can also induce transitions
between different hyperfine multiplets ($F'\neq F$ or $f'\neq f$).
Since the hyperfine multiplets decay rapidly (see Eq. \ref{eq:hyperfine eigenvalues})
in the rapid spin-exchange regime ($R_{SE}\gg\omega_{0}$), all transitions
involving the hyperfine multiplets will be negligible. Even if the
hyperfine multiplets were not decaying rapidly (i.e. in the presence
of applied Rf fields) the rapid hyperfine oscillations would average
the scattering and make it negligible. Thus, the representation of
the super-operator $Q$ by the magnetic set $\left|LM\pm\right\rangle \left\langle lm\pm'\right|$
in the main text is adequate.